\documentclass{PoS}

\usepackage{graphicx}

\usepackage{amsmath}
\usepackage{comment}

\newcommand{\hatbnu}{\widehat{\boldsymbol {\nu}}}
\newcommand{\vn}{ {\bf n} }

\newcommand{\cAb}{{\overline{\cal A}}}
\newcommand{\cFb}{{\overline{\cal F}}}
\newcommand{\cDb}{{\overline{\cal D}}}

\newcommand{\Tr}{\;{\rm Tr}}
\newcommand{\hf}{\frac{1}{2}}

\def\phib{{\overline{\phi}}}

\def\etab{{\overline{\eta}}}
\def\psib{{\overline{\psi}}}

\def\nn{\nonumber}

\newcommand{\cA}{{\cal A}}
\newcommand{\cD}{{\cal D}}
\newcommand{\cF}{{\cal F}}
\newcommand{\cN}{{\cal N}}
\newcommand{\cQ}{{\cal Q}}
\newcommand{\cU}{{\cal U}}

\newcommand{\rR}{{\rm R}}
\newcommand{\rRb}{\bar{{\rm R}}}

\def\phib{{\bar{\phi}}}

\def\etab{{\bar{\eta}}}
\def\psib{{\bar{\psi}}}
\def\chib{{\bar{\chi}}}

\def\nn{\nonumber}

\def\bec{\begin{center}}
\def\eec{\end{center}}
\def\beq{\begin{equation}}
\def\eeq{\end{equation}}
\def\bea{\begin{eqnarray}}
\def\eea{\end{eqnarray}}

\title{\vspace{-3.5cm}{\small \normalfont \hfill DESY-14-173}\vspace{3cm}\\ Lattice formulations of supersymmetric gauge theories with matter fields}

\ShortTitle{Lattice formulations of supersymmetric gauge theories with matter fields}

\author{\speaker{Anosh Joseph} \\
John von Neumann Institute for Computing (NIC), DESY, Platanenallee 6, D-15738 Zeuthen, Germany \\
E-mail: \email{anosh.joseph@desy.de}}

\abstract{Certain classes of supersymmetric gauge theories, including the well known ${\mathcal N} = 4$ supersymmetric Yang-Mills theory, that takes part in the AdS/CFT correspondence, can be formulated on a Euclidean spacetime lattice using the techniques of exact lattice supersymmetry. Great ideas such as topological field theories, Dirac-K\"ahler fermions, geometric discretization all come together to create supersymmetric lattice theories that are gauge-invariant, doubler free, local and exact supersymmetric. We discuss the recent lattice constructions of supersymmetric Yang-Mills theories in two and three dimensions coupled to matter fields in various representations of the color group.}

\FullConference{The 32nd International Symposium on Lattice Field Theory,\\23-28 June, 2014\\Columbia University, New York, NY}

\begin{document}

\section{Introduction}
\label{sec:intro}

Supersymmetric field theories form interesting classes of theories by themselves. Many phenomenologically relevant models can be constructed using these theories as starting points. Supersymmetric field theories generally come with strongly interacting sectors that are less tractable analytically. If one could formulate them on a Euclidean spacetime lattice, in a consistent manner, that would serve as a first-principles approach to explore the nonperturbative regimes of those theories. Here we discuss the exact supersymmetric lattice constructions of certain classes of supersymmetric field theories, namely supersymmetric Yang-Mills (SYM) theories with extended supersymmetries coupled to matter fields. There are two approaches readily available to us for the lattice constructions of SYM theories preserving at least one supercharge exact on the lattice. They are called the methods of topological twisting and orbifolding. They both give rise to supersymmetric lattices that are identical \cite{Giedt:2006pd}.

Supersymmetric lattices have been constructed for several classes of SYM theories in the recent past including the well known $\cN = 4$ SYM in four dimensions \cite{Catterall:2005fd}. There have been a few extensions of these formulations by adding matter in the adjoint and fundamental representations of the gauge group \cite{Endres:2006ic, Matsuura:2008cfa, Joseph:2013jya}. Some of them have also been extended to include product gauge groups, resulting in supersymmetric quiver gauge theories on the lattice \cite{Matsuura:2008cfa, Joseph:2013jya, Joseph:2013bra, Joseph:2014bwa}. 

Here we detail the lattice constructions of two- and three-dimensional SYM theories, with four and eight supercharges, coupled to matter fields transforming in the fundamental or two-index representations of SU($N_c$) color group. These theories are constructed using the following general procedure. We begin with a Euclidean SYM on ${\mathbb R}^d$ possessing $Q = 2^d$ number of supercharges. The fields and supersymmetries of this theory are then topologically twisted to obtain a continuum theory, which is compatible with lattice discretization. This theory is then dimensionally reduced to ${\mathbb R}^{(d-1)}$. The resultant theory now contains matter in the adjoint representation. The next step is to construct a supersymmetric quiver gauge theory with two nodes and with gauge group SU($N_c$) $\times$ SU($N_f$). This can be achieved by making two copies of the continuum twisted theory on ${\mathbb R}^{(d-1)}$, and then changing the gauge group representation of an appropriate subset of the field content of the theory from adjoint to the product representations $(N_c, \overline{N}_f)$ and $(\overline{N}_c, N_f)$, with $N_c$ and $N_f$ being the fundamental representations of SU($N_c$) and SU($N_f$), respectively. At this point we note that we could also change the matter representation of the quiver theory to two-index representation of SU($N$), which we denote by the product representations $(\rR, \rRb)$ and $(\rRb, \rR)$, with $\rR$ the desired representation and $\rRb$ the corresponding complex conjugate representation. The adjoint fields of the quiver theory live on the nodes while the fields in the product representations live on the links connecting the nodes. To construct two- or three-dimensional lattice gauge theories with fundamental or two-index matter, we freeze the theory on one of the nodes of the quiver and also an appropriate set of matter fields linking the two nodes. After this restriction, we have an SU($N_c$) gauge theory on ${\mathbb R}^{(d-1)}$ containing matter in the fundamental or two-index representation and with SU($N_f$) flavor symmetry. Note that such a restriction of the fields is not in conflict with supersymmetry. The continuum theories constructed this way, with fundamental or two-index matter, can then be placed on the lattice using the method of geometric discretization. The resultant lattice theories contain adjoint fields that form components of Dirac-K\"ahler field, and take assignments and orientations on the p-cells of a $(d-1)$-dimensional hypercubic lattice. The matter fields of the theory occupy the sites of the lattice. The lattice theories constructed this way are gauge-invariant, doubler free and exact-supersymmetric at finite lattice spacing. We note that such theories could also be obtained from appropriate orbifold projections of suitable ``parent'' theories but we will not take that path of construction here. 

\section{Two-dimensional $\cN=(2, 2)$ theories with matter}
\label{sec:2dsym}

We begin with writing down the topologically twisted action of the two-dimensional Euclidean SYM with eight supercharges. The key idea of topological twisting is to decompose the fields and supersymmetries of the theory in terms of representations of a newly defined rotation group. The new rotation group is the diagonal subgroup of the product of the two-dimensional Euclidean Lorentz rotation group SO($2$)$_{\rm E}$ and the SO($2$) subgroup of the R-symmetry group of the original theory. (Here we use the so-called B-model twist.) After topological twisting, the fermions of the twisted theory become p-forms with p$=0, 1, 2$ and they are labeled as $\{\eta$, $\psi_m$, $\chi_{mn}$, $\etab$, $\psib_m$, $\chib_{mn}\}$, with $m, n = 1, 2$.  The twisted theory contains two vector fields - the gauge field $A_m$ and a field $B_m$ composed of two scalars of the untwisted theory. We can combine them to form a complexified gauge field $\cA_m = A_m + iB_m$. The theory also contains two scalars $\phi$ and $\phib$. All fields take values in the adjoint representation of the gauge group.

The supercharges of the original theory undergo a decomposition similar to that of the fermions after twisting. They are denoted by $\cQ$, $\cQ_m$, $\cQ_{mn}$ and are called the twisted supercharges. The twisted supersymmetry algebra can be written as: $\cQ^2 = 0,~\{\cQ, \cQ_m\} = p_m,~...~.$ These equations imply that we can construct a lattice action in a $\cQ$-exact form and it is trivially invariant under the (scalar) supercharge $\cQ$. Thus the process of twisting can be used to construct a lattice action that preserves at least one supersymmetry exact on the lattice. The lattice theories constructed using twisted fermions are also free from the fermion doubling problem owing to their geometric nature (they are p-forms) and thus can be mapped one-to-one on to the lattice from continuum.

The action of the two-dimensional theory can be written as 
\beq
\label{eq:2d-adj-matter}
S = S_{{\rm SYM}}^{\cN=(2, 2)} + S_{{\rm adj~matter}},
\eeq
where the first piece is the action of the two-dimensional $\cN=(2, 2)$ SYM
\beq
\label{eq:2d_Q4_action}
S_{{\rm SYM}}^{\cN=(2, 2)} = \frac{1}{g_2^2} \int d^2x \Tr \Big(-\cFb_{mn} \cF_{mn} + \frac{1}{2}[\cDb_m, \cD_m]^2 - \chi_{mn} \cD_{[m} \psi_{n]} - \eta \cDb_m \psi_m \Big),
\eeq
with $g_2$ the coupling parameter of the theory, and the second piece has the form
\bea
S_{{\rm adj~matter}} &=& \frac{1}{g_2^2} \int d^2 x~\Tr \Big(-2(\cDb_m\phib)(\cD_m\phi) + [\cDb_m, \cD_m][\phib, \phi] + \etab \cD_m \psib_m \nn \\
&&+ \chib_{mn}\cDb_{[m} \psib_{n]} - \eta [\phib, \etab] - \psi_m [\phi, \psib_m] - \hf \chi_{mn} [\phib, \chib_{mn}] + \hf [\phib, \phi]^2 \Big).
\eea

The theory contains complexified covariant derivatives and they are defined by $\cD_m ~\cdot = \partial_m \cdot + ~[\cA_m, ~\cdot~ ]$ and $\cDb_m ~\cdot = \partial_m \cdot + ~[\cAb_m, ~\cdot~ ]$. The complexification of gauge field also results in complexified field strength $\cF_{mn} = [\cD_m, \cD_n]$ and $\cFb_{mn} = [\cDb_m, \cDb_n]$.  

The scalar supersymmetry acts on the fields the following way: $\cQ \cA_m = \psi_m,~\cQ \psi_m = 0,~\cQ \cAb_m = 0,~\cQ \chi_{mn} = -\cFb_{mn},~\cQ \eta = d,~\cQ d = 0$, with $d$ an auxiliary field with the equation of motion $d = \sum_m [\cDb_m, \cD_m]$.

\subsection{$\cN = (2, 2)$ theories with two-index matter}
\label{sec:sym_2d_2index_matter}

We are interested in constructing two-dimensional $\cN = (2, 2)$ lattice gauge theories coupled to matter fields transforming in the two-index representations of SU($N_c$) gauge group. To construct such theories we follow the procedure detailed in Refs. \cite{Joseph:2013jya, Joseph:2014bwa}. 

\begin{figure}
\begin{center}
\includegraphics[width=0.35\textwidth]{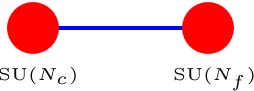}
\end{center}
\caption{\label{fig:quiver_diagram}A quiver diagram for SU($N_c$) $\times$ SU($N_f$) gauge theory. A field in the product representation live on the link connecting the two nodes.}
\end{figure}

To construct the action with two-index matter we note that we can rewrite the action given in Eq. (\ref{eq:2d-adj-matter}) such that the theory becomes a two-dimensional quiver gauge theory with $\cN=(2, 2)$ supersymmetry. In the case we are interested in, there are two interacting SU($N$) gauge theories. The SYM multiplets of this quiver gauge theory transform in the adjoint representation of the gauge group SU($N_c$) $\times$ SU($N_f$) and the two theories interact via matter multiplets in the bi-fundamental representation of the product gauge group. (See Fig. \ref{fig:quiver_diagram}.) The action of the quiver theory can be decomposed in the following way
\beq
S = S^{\rm SYM}_{({\bf adj}, {\bf 1})} + S^{\rm SYM}_{({\bf 1}, {\bf adj})} + S^{\rm matter}_{(N_c, \overline{N}_f)} + S^{\rm matter}_{(\overline{N}_c, N_f)},
\eeq
with the field content of the theory $\{\cA_m$, $\cAb_m$, $\eta$, $\psi_m$, $\chi_{mn}\}$, $\{\widehat{\cA}_m$, $\widehat{\cAb}_m$, $\widehat{\eta}$, $\widehat{\psi}_m$, $\widehat{\chi}_{mn}\}$, $\{\phi$, $\widehat{\phib}$, $\etab$, $\widehat{\psib}_m$, $\chib_{mn}\}$ and $\{\widehat{\phi}$, $\phib$, $\widehat{\etab}$, $\psib_m$, $\widehat{\chib}_{mn}\}$ transforming respectively as $({\bf adj}, {\bf 1})$, $({\bf 1}, {\bf adj})$, $(N_c, \overline{N}_f)$ and $(\overline{N}_c, N_f)$ under SU($N_c$) $\times$ SU($N_f$). The two-dimensional quiver gauge theory with $\cN = (2,2)$ supersymmetry constructed here is interesting in its own right. This quiver theory construction can be easily transported on to the lattice. 

Here we consider matter fields in the two-index representations of SU($N$) transforming as $(\rR, \rRb)$ and $(\rRb, \rR)$ under the product gauge group, with $\rR$ being the desired two-index representation and $\rRb$ the corresponding complex conjugate representation. One could, in principle, look for a suitable projection method that leads to matter in the two-index representation but here we impose such a requirement {\it by hand}. The matter fields of the quiver theory live on the links connecting the two nodes of the quiver. The next step is to freeze one of the nodes of the theory, say the $N_f$-node and make one set of the link fields non-dynamical by hand. The resulting theory is a two-dimensional $\cN = (2, 2)$ gauge theory with matter in the two-index representation of SU($N_c$) gauge group and with SU($N_f$) flavor symmetry.

The action of the continuum theory with two-index matter has the form $S = S_{{\rm SYM}}^{\cN=(2, 2)} + S_{{\rm 2I~matter}}$, where the first part is the same as the one given in Eq. (\ref{eq:2d_Q4_action}). The matter part of the action is given by 
\bea
S_{{\rm 2I~matter}} &=& \frac{1}{g_2^2} \int d^2 x~\Tr \Big(2 \phib^\alpha (\cDb_m\cD_m) \phi^\alpha - [\cDb_m, \cD_m] (\phi^\alpha \phib^\alpha) + \psib^\alpha_m \cD_m \etab^\alpha + \chib^\alpha_{mn}\cDb_{[m} \psib^\alpha_{n]} \nn \\
&&+ \eta \etab^\alpha \phib^\alpha - \psi_m \phi^\alpha \psib^\alpha_m + \hf \chi_{mn} \chib^\alpha_{mn} \phib^\alpha + \hf (\phi^\alpha \phib^\alpha)^2 + \hf (\phib^\alpha \phi^\alpha)^2 \Big),~~~~
\eea
with $\alpha$ an index labeling the SU($N_f$) flavor symmetry.

\section{$\cN = (2, 2)$ lattice theory with two-index matter}
\label{sec:lattice_theories}

\begin{figure}
\begin{center}
\includegraphics[width=0.4\textwidth]{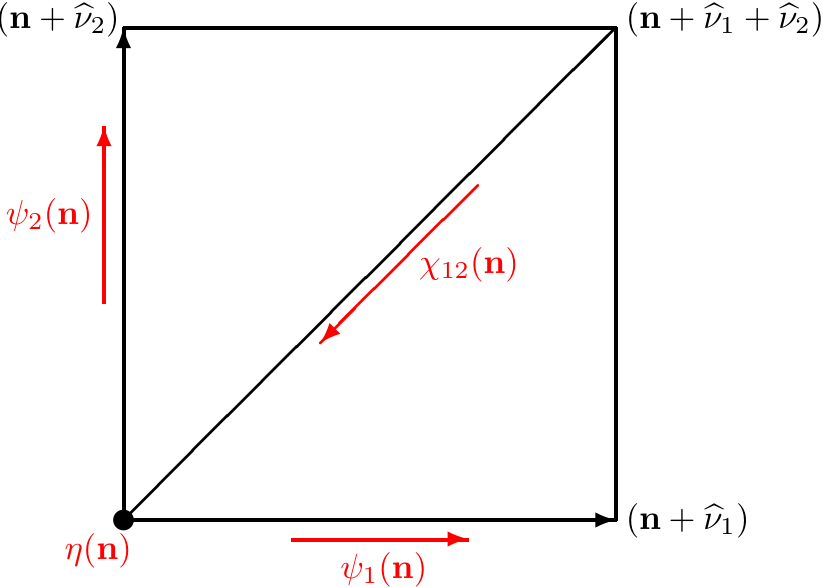}\includegraphics[width=0.38\textwidth]{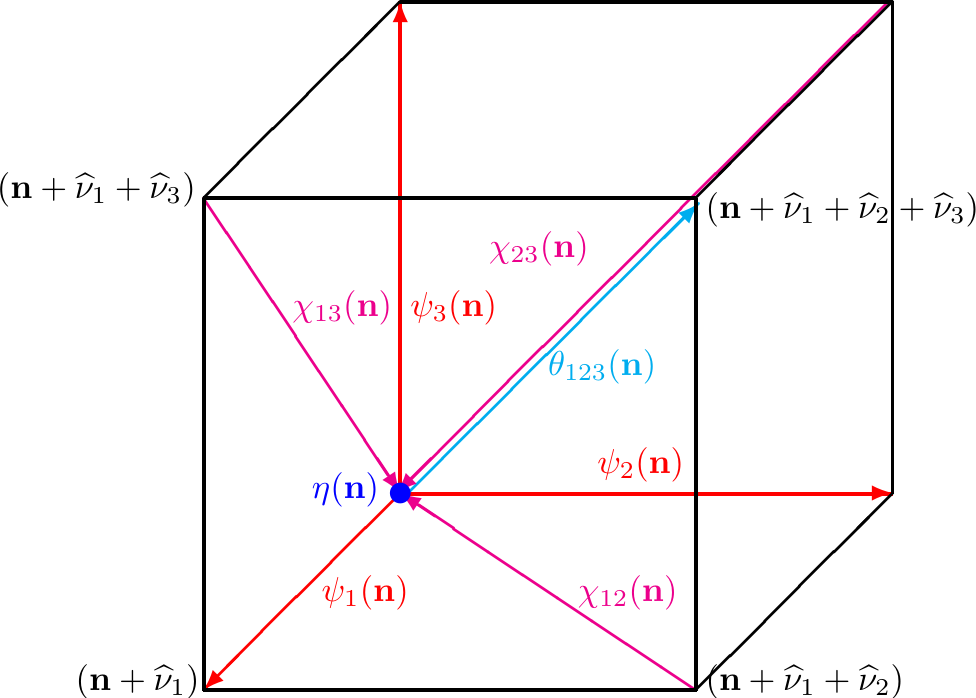}
\end{center}
\caption{\label{fig:2d-3d-latt} (Left) The unit cell of the two-dimensional $\cN = (2, 2)$ lattice SYM with orientation assignments for the twisted fermions. The complexified bosons $\cU_m(\vn)$ follow the same orientations and link assignments as that of their superpartners $\psi_m(\vn)$. The lattice field strength $\cF_{mn}(\vn)$ is placed on the diagonal link with the orientation opposite to that of the field $\chi_{mn}(\vn)$. (Right) The unit cell of three-dimensional $\cN=4$ lattice SYM with orientation assignments for the twisted fermionic fields.}
\end{figure}

We use the technique of geometric discretization \cite{Damgaard:2007be} to write down the lattice theory. We replace the continuum complex gauge field $\cA_m(x)$, by an appropriate complexified Wilson link $\cU_m(\vn)$. These lattice fields are taken to be associated with unit length vectors in the coordinate directions $\hatbnu_m$ from the site denoted by the integer vector $\vn$ on a two-dimensional square lattice. The unit cell and the field orientations of the two-dimensional theory are given in Fig. \ref{fig:2d-3d-latt}.
The covariant derivatives $\cD_a$ ($\cDb_a$) in the continuum become forward and backward covariant differences $\cD^{(+)}_a~(\cDb^{(+)}_a)$ and $\cD^{(-)}_a~(\cDb^{(-)}_a)$, respectively. As an example, we have the following rule for the action of $\cD^{(+)}_a$ on the positively oriented field $f^{(+)}_{a_1 \cdots a_p}(\vn)$: $\cD_b^{(+)}f^{(+)}_{a_1 \cdots a_p}(\vn) \equiv \cU_b(\vn)f^{(+)}_{a_1 \cdots a_p}(\vn + \hatbnu_b)-f^{(+)}_{a_1 \cdots a_p}(\vn) \cU_b(\vn+\hatbnu)$ with $\hatbnu = \sum_{i=1}^p \hatbnu_{a_i}$. We also need to specify the rules for the action of the covariant difference operators on the lattice fields transforming in the two-index representations. The rule for the action of $\cD^{(+)}_a$ on the lattice variables in the two-index representation is given by: $\cD_m^{(+)} \Phi^{\rR}(\vn) \equiv \cU_m(\vn) \Phi^{\rR}(\vn + \hatbnu_m) - \Phi^{\rR}(\vn)$. 

Once we have the rules to map the continuum twisted theory to the lattice it is easy to write down the lattice action. The two-dimensional $\cN = (2, 2)$ gauge theory with two-index matter has the form on the lattice: $S = S_{{\rm SYM}}^{\cN=(2, 2)} + S_{\rm 2I~matter}$, with
\bea
S_{{\rm SYM}}^{\cN=(2, 2)} &=& \frac{1}{g_2^2}\sum_{\vn} \Tr~ \Big(-\cFb_{mn}(\vn)\cF_{mn}(\vn) + \frac{1}{2}\Big(\cDb_m^{(-)}\cU_m(\vn)\Big)^2 -\chi_{mn}(\vn)\cD_{[m}^{(+)}\psi_{n]}(\vn) \nn \\
&&- \eta(\vn) \cDb_m^{(-)}\psi_m(\vn)\Big)
\eea
and
\bea
S_{{\rm 2I~matter}} &=& \frac{1}{g_2^2} \sum_{\vn} \Tr \Big(2 \phib^\alpha(\vn) \cDb^{(-)}_m \cD^{(+)}_m \phi^\alpha(\vn) - \Big(\cDb^{(-)}_m \cU_m(\vn)\Big) \Big(\phi^\alpha(\vn) \phib^\alpha(\vn)\Big) \nn \\
&&+ \psib^\alpha_m(\vn) \cD^{(+)}_m \etab^\alpha(\vn) + \chib^\alpha_{mn}(\vn) \cDb^{(+)}_m \psib^\alpha_n(\vn) + \eta(\vn) \etab^\alpha(\vn) \phib^\alpha(\vn) \nn \\
&&- \psi_m(\vn) \phi^\alpha(\vn + \hatbnu_m) \psib^\alpha_m(\vn) + \hf \chi_{mn}(\vn) \chib^\alpha_{mn}(\vn) \phib^\alpha(\vn + \hatbnu_m + \hatbnu_n) \nn \\
&&+ \hf (\phi^\alpha(\vn) \phib^\alpha(\vn))^2 + \hf (\phib^\alpha(\vn) \phi^\alpha(\vn))^2 \Big).
\eea
 
We note that the lattice action written above is $\cQ$-supersymmetric, gauge-invariant, local and free from the fermion doublers. 

\section{Three-dimensional $\cN=4$ SYM with matter}
\label{sec:3dsym}

The twisted rotation group of the three-dimensional $\cN = 4$ SYM is defined as $SU(2)' = {\rm diag}\left(SU(2)_E \times SU(2)_N\right)$. After twisting, the theory contains a three-dimensional gauge field $A_m$, $m=1,2,3$; a vector $B_m$ composed of three scalars of the untwisted theory; and eight p-form fermions, p $= 0, 1, 2, 3$, which we conveniently represent as $\{\eta, \psi_m, \chi_{mn}, \theta_{mnr}\}$.  Here also we can combine the two vector fields to obtain a complexified gauge field.

The nilpotent scalar supersymmetry $\cQ$ acts on the fields the following way: $\cQ \cA_m = \psi_m$,  $\cQ \psi_m = 0$, $\cQ \eta = d$,  $\cQ B_{mnr} = \theta_{mnr}$,  $\cQ \cAb_m = 0$, $\cQ \chi_{mn} = -\cFb_{mn}$,  $\cQ d = 0$, $\cQ \theta_{mnr} = 0$, where $B_{mnr}$ and $d$ are auxiliary fields.

We can add fundamental matter to the three-dimensional $\cN = 4$ theory by following the construction procedure mentioned above. The action of the theory with fundamental matter can be written as: $S = S^{\rm SYM} + S^{\rm matter}$, where $S^{\rm SYM}$ contains adjoint fields
\bea
\label{eq:adj-3d-sym}
S^{\rm SYM} &=& \frac{1}{g_3^2}\int d^3 x~\Tr \Big(-\cFb_{mn}\cF_{mn} + \hf[\cDb_m, \cD_m]^2 -\chi_{mn} \cD_{[m}\psi_{n]} - \psi_m\cDb_m\eta - \theta_{mnr}\cDb_{\left[r \right.} \chi_{\left.mn\right]}\Big),~~~~~~
\eea
with $g_3$ the coupling parameter of the theory. The piece $S^{\rm matter}$ contains matter fields in the fundamental representation. (See Ref. \cite{Joseph:2013jya} for the detailed form of the matter action.)

\section{Lattice formulation of three-dimensional $\cN=4$ SYM with fundamental matter}
\label{sec:latt-form-3dSYM}

Having the prescription for geometric discretization in hand, we can write down the supersymmetric and gauge-invariant lattice action of the three-dimensional $\cN = 4$ SYM with fundamental matter. The SYM part of the action, given in Eq. (\ref{eq:adj-3d-sym}), takes the form
\bea
\label{eq:3d-latt-action}
S^{\rm SYM} &=& \frac{1}{g_3^2}\sum_{\vn} \Tr~ \Big(-\cFb_{mn}(\vn)\cF_{mn}(\vn) + \frac{1}{2}\Big(\cDb_m^{(-)}\cU_m(\vn)\Big)^2 -\chi_{mn}(\vn)\cD_{[m}^{(+)}\psi_{n]}(\vn) \nn \\
&&- \eta(\vn) \cDb_m^{(-)}\psi_m(\vn) - \theta_{mnr}(\vn)\cDb_{[r}^{(+)}\chi_{mn]}(\vn)\Big),
\eea
with another piece of the action, the discretized form of $S^{\rm matter}$, containing the fundamental matter with appropriate placements for the lattice fields on a cubic lattice. The unit cell of the lattice theory is given in Fig. \ref{fig:2d-3d-latt}. We need to specify the rule for the action of the covariant difference operators on fundamental fields. As an example, we have the rule for the action of $\cD_b^{(+)}$ on the field $f^{(+)\Box}_{a_1 \cdots a_p}$: $\cD_b^{(+)}f^{(+)\Box}_{a_1 \cdots a_p}(\vn) = \cU_b(\vn) f^{(+)\Box}_{a_1 \cdots a_p}(\vn + \hatbnu_b)$.

\section{Discussion and comments}
\label{sec:discussion_comments}

We have detailed the continuum and lattice constructions of two- and three-dimensional SYM theories with matter in the fundamental and two-index representations of the color group. The lattice theories constructed this way are gauge-invariant, free from fermion doublers and preserve at least one supersymmetry exact on the lattice. The lattice theories discussed here would be interesting in the context of string theory. Three-dimensional $\cN = 4$ gauge theories play an important role in our understanding of dualities. It would be interesting to explore the two-dimensional lattice theories with matter discussed here in the context of the Corrigan-Ramond limit. In string theory, one can obtain the two-index representation of SU($N_c$) by performing the orientifold projection to adjoint of SO($2N_c$) or Sp($2N_c$). The continuum (and thus lattice) theories we considered here would be obtained in the same manner.



{\bf Acknowledgments:} This work was supported in part by the EU-project Hadron Physics III.

\end{document}